\documentclass[journal]{IEEEtran}

\usepackage{amsmath,nccmath}
\usepackage{graphicx}
\usepackage{float}
\usepackage{cite}
\begin{document}
\setlength{\abovedisplayskip}{1pt}
\setlength{\belowdisplayskip}{1pt}
\setlength{\abovecaptionskip}{-3pt}

\title{Revenue Analysis of Stationary and Transportable Battery Storage for Power Systems with High Penetration of Renewable Sources: A Market Participant Perspective}

\author{Zhongyang~Zhao,~\IEEEmembership{Student~Member,~IEEE,}
        Caisheng~Wang,~\IEEEmembership{Senior~Member,~IEEE,}
        and~Masoud~H.~Nazari,~\IEEEmembership{Senior~Member,~IEEE}\vspace{-3ex}
\thanks{The authors are with the Department
of Electrical and Computer Engineering, Wayne State University, Detroit,
MI, 48202 USA (e-mail: zhongyang.zhao@wayne.edu, cwang@wayne.edu,
masoud.nazari@wayne.edu).}
}


%

\maketitle

\begin{abstract}
The power system needs more transmission capacity to deal with the increasing integration of renewable energy resources. The battery energy storage systems (BESSs) are effective to enhance the grid capacity and relieve the transmission congestion. A comprehensive revenue analysis of BESSs is critical for market participants to install such systems in a market-based power system. Taking PJM as an example, this paper carries out a thorough revenue analysis for the entire system. Highly profitable nodes in the system are revealed and characterized for BESS participants. A comparison study of stationary and transportable BESSs shows the transportable energy storage can produce higher potential revenue in the energy and regulation markets. Based on the results of the revenue analysis and characterization of commercial pricing nodes, an optimal placement algorithm is proposed for finding the profitable sites for market participants to install BESSs in the system and the algorithm is validated with real PJM market data.
\end{abstract}


%
\IEEEpeerreviewmaketitle

\section{Introduction}
%
\IEEEPARstart{F}acing the challenges of climate crisis, several states in the U.S. have announced their targets to achieve the net-zero emission by 2050 \cite{net_zero_state}. As the electric power sector shares 33\% of carbon dioxide emission in the U.S., the emission reduction of the power generation holds a key to reach the net-zero emission \cite{emission_by_sector}. Using the renewable energy resources helps to reduce greenhouse gas emissions. Since 2018, the generation capacity of wind and solar has increased over 30\% and 65\%, respectively. However, the integration of renewable generation has challenged the operation of power system. Due to a lack of transmission capacity, the renewable generation has caused the transmission congestion and needs to be significantly curtailed \cite{berizzi2015decentralized, schermeyer2017understanding, SPP_wind_curtailment}. Furthermore, numbers of renewable energy projects have to be withdrawn from the interconnection queue in some highly congested regions \cite{miso_renewable_withdraw} while building the transmission lines is costly. \par
Under this situation, the role of battery energy storage systems (BESSs) becomes more and more important as they can provide a variety of functions \cite{sioshansi2012market}, including energy arbitrage, ancillary services, generation capacity deferral, ramping, congestion relief and accommodation of intermittent renewable energy. To benefit the operations of power systems from the implementation of BESS, Federal Energy Regulatory Commission (FERC) has issued Orders 755 \cite{Order755} and 841 \cite{Order841} in 2011 and 2018, respectively, which have further paved the way for market participation of energy storage. Order 755, titled “Frequency Regulation Compensation in the Organized Wholesale Power Markets,” requires the Independent System Operators (ISOs) and Regional Transmission Organizations (RTOs) to pay for the capacity offers and regulation performances provided by energy storage. Order 841, called “Electric Storage Participant in Markets Operated by Regional Transmission Organizations and Independent System Operators,” guides ISOs and RTOs toward lowering the barriers for the participation of electric storage resources in the capacity, energy, and ancillary service markets \cite{Order841}. As the battery technologies advance and the cost of battery decreases, BESSs have emerged as one of the popular and cost-effective energy storage technologies for power system applications. Recently, the utility-scale BESS applications have experienced exponential growth worldwide. For example, the BESS capacity in the U.S. has increased from 845.5 MW in 2018 to 2915.9 MW in 2021 \cite{EIA-860M}. \par
In addition, the integration of renewable energy sources has caused serious transmission congestion in the power markets, which makes the Locational Marginal Prices (LMPs) more volatile in the congested regions. The increasing arbitrage opportunities will encourage more BESSs to participate in the wholesale market. Meanwhile, BESSs can also quickly respond to the system control signals, provide mobile and highly flexible storage capacity, and can be placed at desirable locations in the grid for optimal operations, which is able to improve the power system reliability and better address the energy crisis in the future. \par
%
%
However, having access to the wholesale market and the new profit opportunities do not mean to arbitrarily build more BESS projects to dive into the market while many factors can influence the profits. Due to the volatility and variations of LMP, Regulation Market Clearing Price (MCP), and automatic generation control (AGC) regulation signal of different locations and seasons in the power market, the investment return for market participants can be vastly different. A poor selection of installation location for a BESS can make it struggle. Therefore, a comprehensive analysis for finding the locations with the best potential revenues and mitigate the transmission congestion in the wholesale markets is a high-priority task for the market participants who plan to invest in BESS. To analyze the potential income of BESS in the energy and regulation markets, several studies have been done \cite{xu2016comparison,wang2016two,SandiaStorageReport,MISO2017maximizing,PJM2016estimating,ERCOT2015potential,CAISO2018opportunities,NewYork2007economics}. The energy storage technologies, potential applications, and comparison of policies on the participation of BESS in different markets were investigated and studied in \cite{xu2016comparison,wang2016two,SandiaStorageReport}. The analyses of potential revenues have been carried out in MISO \cite{MISO2017maximizing}, PJM \cite{PJM2016estimating}, CAISO \cite{CAISO2018opportunities}, ERCOT \cite{ERCOT2015potential}, and New York ISO \cite{NewYork2007economics}. Although the potential revenues of BESS in energy and frequency regulation markets were estimated in \cite{MISO2017maximizing, PJM2016estimating}, the estimations were only based on a single node, such as IPL.16STOU6O6 in MISO \cite{MISO2017maximizing} and HAZLETON 1-4 in PJM \cite{PJM2016estimating}. In \cite{ERCOT2015potential}, the possible revenues for BESS were estimated only for the load zones of ERCOT.  Because of the insufficient study from the perspective of the whole system, the characteristics of the potential revenues throughout the entire power market have not been revealed for the BESS participants yet. \par
%
%
In addition to the stationary system, BESS can be installed on a mobile platform such as a truck or an aggregation of EVs with Vehicle to Grid (V2G) capability. While a transportable BESS can be implemented at different locations during operation, it is also imperative for the market participants to have an estimation on the current and future status of the power market and deploy the BESS at the appropriate site to better accommodate to the renewable energy resources and transmission congestion. Therefore, in addition to the temporal system-wide revenue estimation, it is beneficial to optimally locate the portable BESSs to maximize their profitability. However, most of the existing studies on portable BESSs focus on the operation of power systems for system operator, such as load shifting \cite{khodayar2013electric, knezovic2016enhancing}, transmission congestion relief \cite{sun2015battery, he2018spatiotemporal}, enhancing the resilience of distribution system \cite{yan2017lmp, yao2018transportable}, and the reduction of wind curtailment \cite{sun2016stochastic}. An optimal placement algorithm to maximize revenue in energy and frequency regulation markets from the perspective of a market participant has not been fully investigated yet.\par%

%
Therefore, the main contributions of this paper can be summarized as follows: 1) Considering LMPs, regulation MCP and AGC signals, and battery degradation, the potential revenues for BESSs to participate in both the energy and regulation markets are estimated for every node, and the features of nodal profit differences are characterized. 2) The temporal and spatial revenue analysis is performed by considering transportable BESSs to grasp more profitable opportunities. 3) According to the characteristics discovered at the most profitable node, an optimal placement algorithm is proposed for market participants to find the profitable locations for better installation and operation of the BESS, particularly when it is transportable. \par
%
The rest of this paper is organized as follows: Section \ref{ParticipationModels} introduces the participation models for estimating the potential revenues of BESS in energy and regulation markets. Section \ref{MarketRevenue} provides the numerical results and detailed revenue analysis of different cases. An optimal placement algorithm of transportable BESSs is proposed in Section \ref{RevenuePrediction}. The conclusion is drawn in Section \ref{conclusion}. \par
%

\section{Participation Models in Energy and Frequency Regulation Markets} \label{ParticipationModels}
\vspace{-0.1cm}
The following models for participating in PJM energy and frequency regulation markets are used to estimate the potential profit at each commercial pricing node (CPN), which is valid for auction. It is worth noting that, unlike the system operator, the market participants do not have sufficient data to run an accurate DC optimal power flow \cite{frank2016introduction} for the large-scale power market. Hence, the available data for market participants, such as price and regulation signals, are used in the following models for their decisions. \par
\vspace{-0.5cm}
\subsection{Credit in Energy Market}
\vspace{-0.1cm}
The arbitrage credit of energy storage from the energy market during a period of time $T$ is calculated by \eqref{Credit_E}. The time interval is assumed to be one hour.
\begin{equation} \label{Credit_E}
{Credit}_{E} = \sum^{T}_{t=1}P^{E,dis}_t\times{LMP}_t-\sum^{T}_{t=1}P^{E,ch}_t\times{LMP}_t
\end{equation}
where $P^{E,ch}_t$ and $P^{E,dis}_t$ represent the charging and discharging power in the energy market at hour $t$; ${LMP}_t$ is the LMP at hour $t$. \par
\vspace{-0.5cm}
\subsection{Credit in Regulation Market}
In addition to the energy market, energy storage can participate in the regulation market. Since this study focuses on analyzing the electric storage projects competing in PJM, the credit calculation method of participating in the frequency regulation market follows the PJM’s rules. According to the section about regulation credits in the PJM manual \cite{PJMManual28}, the regulation remuneration is obtained from two parts: capability and performance, as shown in \eqref{Credit_R}.
\begin{equation}\label{Credit_R} {Credit}_{R}={Credit}_{cap}+{Credit}_{perf} \end{equation}
Given an offered capacity $P^R_t$, Regulation Market Capability Clearing Price ${RMCCP}_t$, and performance score $\rho_t$, the ${Credit}_{cap}$ can be calculated by \eqref{MPRE_eq9}. Performance score $\rho_t$ is used to evaluate how well the resource is following the regulation signal \cite{PJMManual12}.
\begin{equation}\label{MPRE_eq9} {Credit}_{cap}=\sum^{T}_{t=1}{P^R_t\times {RMCCP}_t\times\rho_t} \end{equation} \par
In PJM's frequency regulation market, there are two types of regulation signals generated every two seconds. One is the Regulation D signal ($RegD$), which is a fast and dynamic signal for fast-responding resources such as BESS. The other one is Regulation A signal ($RegA$), a slower signal for conventional resources like hydropower \cite{PJMManual12}.
%
\begin{figure}[H] \vspace{-0.1cm}
	\centering
	\includegraphics[width=2.6 in]{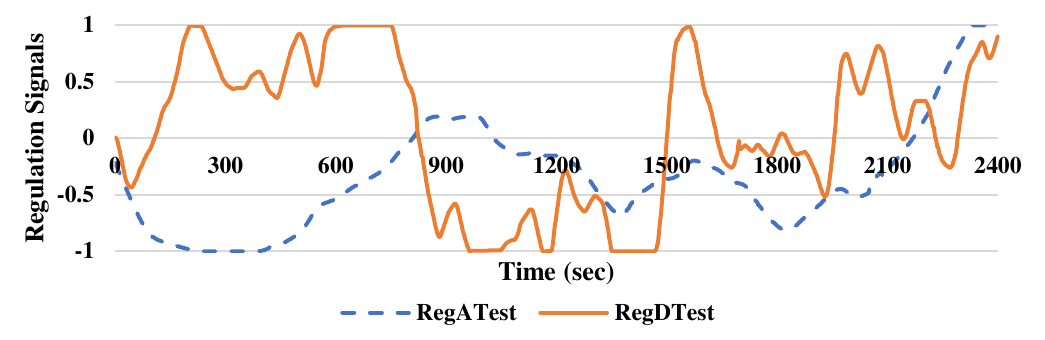}
	\caption{Comparison between $RegA$ and $RegD$.} \vspace{-0.15cm}
	\label{RegA_vs_RegD}\vspace{-0.2cm}
\end{figure} 
To show the difference between these two signals, a series of 40-minute PJM regulation test signals \cite{PJM_Regulation_Test_Signal} are plotted in Fig. \ref{RegA_vs_RegD}. According to the plots, $RegD$ has much higher volatility than $RegA$. The $RegA$ resources typically do not have a duration limit while the $RegD$ resources are expected to have a short period duration \cite{PJM_Regulation_Market_Duration_Limit}. Hence, the BESS is only considered to submit the offer to follow $RegD$. 
In the frequency regulation market, the performance credit ${Credit}_{perf}$ is calculated by \eqref{Credit_perf}. In the equation, $RMPCP_t$ is the Regulation Market Performance Clearing Price at hour $t$ and the Mileage Ratio $\beta_t$ of $RegD$ over $RegA$ can be obtained by \eqref{MPRE_eq11}. The mileage of a regulation signal indicates the total movements of that signal in a given interval. The mileages of $RegD$ and $RegA$ (${Mileage}^{RegD}_t$ and  ${Mileage}^{RegA}_t$ ) are given in \eqref{MPRE_eq121} and \eqref{MPRE_eq122}, respectively. $n$ in those two equations is the number of time intervals. For example, to calculate the mileage of a 2-second signal in one hour, $n$ is 1800.
\begin{gather} \allowdisplaybreaks
{Credit}_{perf}= \sum^{T}_{t=1}{P^R_t\times {RMPCP}_t\times\beta_t\times\rho_t} \label{Credit_perf} \\
\beta_t=\ {Mileage}^{RegD}_t / {Mileage}^{RegA}_t \label{MPRE_eq11} \\
{Mileage}^{RegD}_t= \sum^n_{i=0}{\left|{RegD}_i-{RegD}_{i-1}\right|} \label{MPRE_eq121} 
\end{gather}
\begin{gather}
{Mileage}^{RegA}_t= \sum^n_{i=0}{\left|{RegA}_i-{RegA}_{i-1}\right|} \label{MPRE_eq122} 
\end{gather} \par
\subsection{Cost of Storage Degradation}
\vspace{-0.1cm}
Based on \cite{wang2017improving, wankmuller2017impact}, the degradation cost of BESS is assumed to be a linear function of BESS output as given in \eqref{degradationcost} and \eqref{degradationcost_rate}:
\begin{gather} 
{Cost}_D=\sum^{24}_{t=1}(P^{ch}_t\times\eta_c+P^{dis}_t\times\eta_d^{-1})\times{DEG}_{rate} \label{degradationcost} \\
{DEG}_{rate}=(\upsilon\times\pi_{ES}\times{0.5})/(1-\sigma_{EOL}) \label{degradationcost_rate}
\end{gather}
where $P^{ch}_t$ and $P^{dis}_t$ are the overall charging and discharging power at hour $t$, respectively; the $\eta_c$ and $\eta_d$ are the corresponding charging and discharging efficiency of BESS; ${DEG}_{rate}$ represents the cost of each MWh charge/discharge of the BESS; $\upsilon$ represents the degradation speed of energy storage; $\pi_{ES}$ stands for the cost of energy storage; $\sigma_{EOL}$ means the storage state at the end of life (EOL). \par
\vspace{-0.4cm}
\subsection{Participation of Energy Market}
\vspace{-0.1cm}
When the BESS only participates in the energy market, the objective of energy arbitrage is to maximize the revenue, shown in \eqref{MPRE_ObjectiveEnergy}. We assume that the forecast of the next day electricity price is accurate and the participation model is expected to be price-taking. Therefore, the model for participating in the energy market can be described as: 
\begin{gather}
Max({Credit}_{E}-{Cost}_D) \label{MPRE_ObjectiveEnergy}
\end{gather}
\begin{center}\vspace{-0.2cm} s.t. \eqref{Credit_E}, \eqref{degradationcost}, \eqref{degradationcost_rate}, \eqref{total_output_E1}-\eqref{MPRE_eq5}. \vspace{-0.1cm}\end{center}
\begin{gather}
P^{ch}_t = P^{E,ch}_t \label{total_output_E1}\\
P^{dis}_t = P^{E,dis}_t \label{total_output_E2}\\
0\le P^{E,ch}_t\le P_{max} \label{MPRE_eq21} \\
0\le P^{E,dis}_t \le P_{max} \label{MPRE_eq22} \\
{P^{E,ch}_t}\times{P^{E,dis}_t}=0 \label{MPRE_eqNonlinearConstraint} \\
0\le S_t\le 100\% \label{MPRE_eq3} \\
S_t=S_{t-1}+(P^{ch}_t\times\eta_c-P^{dis}_t\times\eta_d^{-1})/E_{max}\times{100\%} \label{MPRE_eq4} \\
S_0=S_{24} \label{MPRE_eq5}
\end{gather}
where the $P^{ch}_t$ and $P^{dis}_t$ are determined by \eqref{total_output_E1} and \eqref{total_output_E2}. $P_{max}$ and $E_{max}$ are the power rating and energy capacity of the BESS. $S_t$ represents the state of charge (SOC) at time $t$. The operation limits of the BESS are set by \eqref{MPRE_eq21}-\eqref{MPRE_eq3}. $S_0$ and $S_{24}$ are the SOC at $t=0$ and $t=24$ of each day. $S_t$ is obtained from the previous SOC $S_{t-1}$ and the hourly output as given in \eqref{MPRE_eq4}. Equation \eqref{MPRE_eq5} is implemented to ensure that the conditions of daily optimization is consistent. Equation \eqref{MPRE_eqNonlinearConstraint} is used to force the optimization model to not make charging and discharging offers at the same time. Also, \eqref{MPRE_eqNonlinearConstraint} is essential for the market participation model to deal with negative LMPs. \par
%
%
\vspace{-0.3cm}
\subsection{Participation of Energy and Frequency Regulation Markets}
\vspace{-0.1cm}
When the BESS participates in the energy and frequency regulation markets simultaneously, the objective function is formulated by \eqref{MPRE_ObjectiveER} to achieve the maximum daily profit at a node.
\begin{gather}
Max({Credit}_{E}+{Credit}_{R}-{Cost}_D) \label{MPRE_ObjectiveER}
\end{gather}
\begin{center}\vspace{-0.2cm} s.t. \eqref{Credit_E}-\eqref{degradationcost_rate}, \eqref{MPRE_eq21}-\eqref{MPRE_eq5}, \eqref{total_output_ER1}-\eqref{ER_OutputConstraint2}. \vspace{-0.1cm}\end{center}
The $P^{ch}_t$ and $P^{dis}_t$ of the degradation cost for both the energy and regulation markets are calculated by \eqref{total_output_ER1} and \eqref{total_output_ER2}, where ${RegD}^{up}_t$ and ${RegD}^{down}_t$ are the absolute values of hourly accumulations of the regulation up and down signals. In PJM, a unit offering in the regulation market is required to be able to provide the same amount of positive capacity and negative capacity. The constraints of output offers are given in \eqref{R_OutputConstraint}-\eqref{ER_OutputConstraint2}.
\begin{gather}
P^{ch}_t = P^{E,ch}_t+P^R_t\times{RegD}^{down}_t \label{total_output_ER1}\\
P^{dis}_t = P^{E,dis}_t+P^R_t\times{RegD}^{up}_t \label{total_output_ER2}\\
0\le P^R_t\le P_{max} \label{R_OutputConstraint} \\
0\le P^{E,ch}_t+P^R_t\le P_{max} \label{ER_OutputConstraint1} \\
0\le P^{E,dis}_t+P^R_t\le P_{max} \label{ER_OutputConstraint2}
\end{gather} \par
Implementing the introduced participation models with the real market data, such as LMP, RMCCP, RMPCP, and $RegD$, the maximum daily revenue at each node can be obtained and the nodes with the best profitability can be identified, as discussed in Section \ref{MarketRevenue}. 

\vspace{-0.25cm}
\section{Revenue Analysis of BESSs} \label{MarketRevenue}
%
\subsection{Data Preparation} \label{MarketData}
%
There are over 12,000 pricing nodes in PJM, and about 7000 of them are valid for auction \cite{PJM_Valid_Auction_Node}. This paper focuses on analyzing the potential revenue of BESS at the valid nodes. According to the market participation models, introduced in Section II, the LMP, AGC regulation signals, regulation mileage, RMCCP, and RMPCP are required for estimating the potential revenue of participating in the energy and regulation markets. In this paper, the corresponding 12-month market data in 2018 are collected via PJM Data Miner 2 \cite{PJM_Dataminer2}, which is a data management tool to access PJM data. The potential profit of transportable BESS is investigated in different seasons \cite{PJM_Regulation_Requirement_Definition}. \par
%
%
%
In the energy market, LMP reflects the value of electricity at different nodes, which is defined as the cost of supplying the next 1 MW at the node \cite{kirschen2018fundamentals}. LMP is a combination of three components: price of energy, power system congestion, and energy loss. Since these components are affected by load, fuel cost, system topology, etc., the LMPs can have different characteristics and values at different periods and locations, which can cause significant changes in the BESS revenue. 
%
%
%
%
%
Unlike the energy market, the regulation market is pool-based, which implies that there is only one market clearing price for each MW supplied. Hence, all resource owners are credited by the same price for providing the scheduled regulation output. To calculate the potential profit of participating in the regulation market, the hourly LMP, RMCCP, RMPCP, $RegA$ mileage, and $RegD$ mileage in 2018 are collected. To determine the hourly regulation output and the hourly SOC, the 2-second regulation signals \cite{PJM_RTO_Regulation_Signal} are also collected and aggregated to the hourly regulation up and down signals.\par
%
%
In this paper, we study the revenue maximization of a 10 MW/10 MWh BESS.
%
The BESS can achieve charge/discharge ($\eta_{c}/\eta_{d}$) efficiency at 95\%. To estimate the degradation cost, the degradation speed of energy storage $\upsilon$, the cost of energy storage $\pi_{ES}$, and the storage state at the end of life $\sigma_{EOL}$ are set at $3\times10^{-5}$, $1\times10^{5}$ \$/MWh, and 0.8, respectively, \cite{wang2017improving, wankmuller2017impact}. In this study, the initial SOC $S_{0}$ of each day is set to be 50\%. Given these BESS parameters, the degradation rate $DEG_{rate}$ in \eqref{degradationcost_rate} can be obtained as 7.5 \$/MWh. This degradation cost indicates the charging action is worthless for the battery when the profit of 1 MWh charge cycle is less than \$15. 

%
%
\vspace{-0.1cm}
\subsection{Potential Revenue of Stationary BESS}
\vspace{-0.1cm}
The first case study is to place the BESS at one Location for an entire Year (1L1Y), which is the most common setup investigated in the previous studies \cite{xu2016comparison,SandiaStorageReport,MISO2017maximizing,PJM2016estimating,ERCOT2015potential,CAISO2018opportunities,NewYork2007economics,WholePJMpotential,krishnamurthy2017energy,DART_Energy2018maximizing}. The proposed optimal placement algorithm finds a profitable location in the system to build the BESS and obtains the credit settled by the corresponding LMP and regulation MCP for a long period. The 1L1Y utilizes the real market data, including the RTLMP, RMCCP, RMPCP, and the regulation signal, to estimate the potential annual revenues at different settlement price nodes in PJM, which provides the valuable information for implementing BESS from the perspective of market participants. \par
\vspace{-0.1cm}
\begin{figure}[H]
	\centering
	\includegraphics[width=2.8 in]{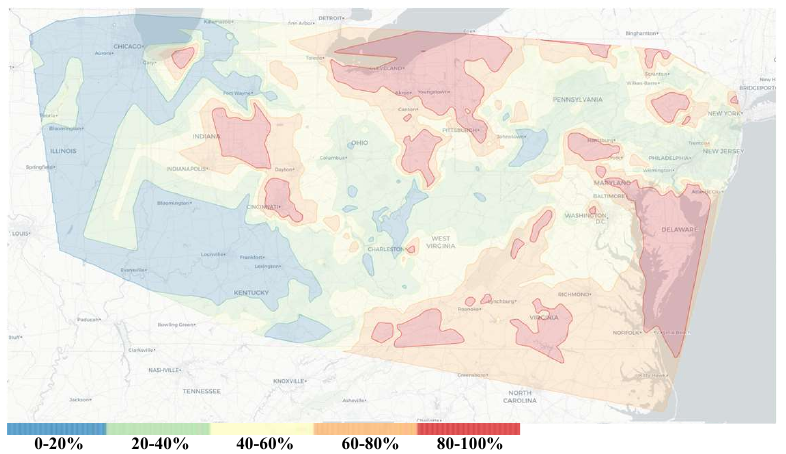}
	\caption{Annual revenue percentiles in PJM.}
	\label{PJM_Annual_ER_RT_Map}
\end{figure}
\begin{figure}
	\centering
	\includegraphics[width=2.8 in]{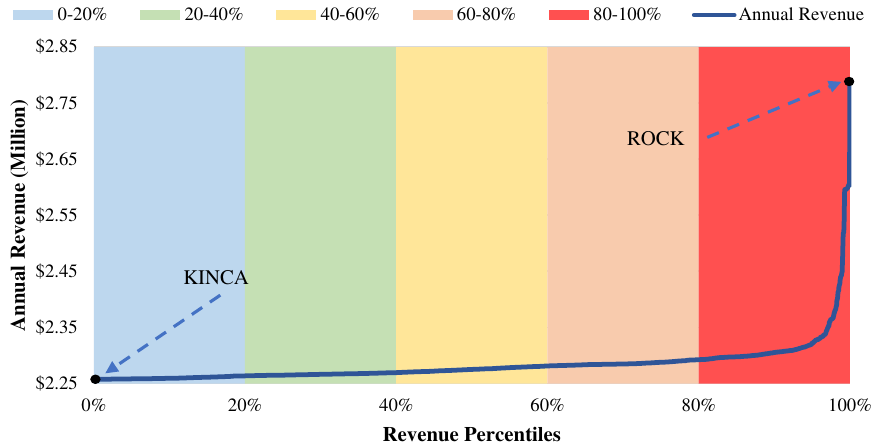}
	\caption{Annual revenues by percentiles.}
	\label{RT_Rev_Percentiles} \vspace{-0.5cm}
\end{figure}

 \vspace{-0.1cm}
Given the 2018 market data collected from PJM, the maximum potential revenue and the corresponding generation outputs of each node can be obtained by the participation models proposed in Section \ref{ParticipationModels}. As shown in Fig. \ref{PJM_Annual_ER_RT_Map}, the percentiles of all nodes' potential revenues are grouped and mapped into five different colors, which indicate the percentiles of 0-20\%, 20-40\%, 40-60\%, 60-80\%, and 80-100\%, respectively. The percentile indicates the revenue's percentage rank of a specified node among all nodes. According to the distributions of the colors on the map, the nodes of the top 20\% (i.e., 80-100\%) annual revenue are found to concentrate in some regions. In other words, the nodes, which have high potential revenue for stationary BESS, are not randomly distributed in the power market. It is feasible for market participants to locate some sites to enhance their competitiveness in the power market. Furthermore, the annual revenues of each node in the five percentile intervals are plotted in Fig. \ref{RT_Rev_Percentiles}, which clearly indicates that there is a sharp increase in revenue for the nodes falling in the 80-100\% percentile interval. Notably, the annual revenue at the most lucrative node, named ROCKWLKN69KV LS-EGRET (ROCK), is \$2.7M, which is \$530k higher than the annual revenue at the least lucrative node, called 21 KINCA20KV KN-1 (KINCA). In other words, the BESS at ROCK could earn 23.5\% more profit than the BESS at KINCA in 2018. \par
\vspace{-0.1cm}
\begin{figure}[H]
	\centering
	\includegraphics[width=3.0 in]{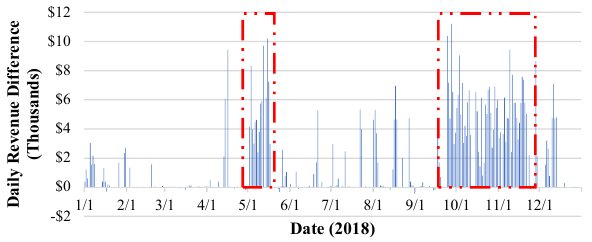} \vspace{-0.15cm}
	\caption{Daily revenue differences between ROCK and KINCA in 2018.}
	\label{Best_Worst_All_Year_Rev_Diff}
\end{figure}

To gain a more comprehensive understanding of the factors causing the revenue disparity, the differences in daily revenues in 2018 between these two nodes are calculated and plotted in Fig. \ref{Best_Worst_All_Year_Rev_Diff}. It can be seen that most of the daily revenue differences are low and the significant revenue differences are in May, October, and November. Moreover, when the daily revenue differences are sorted into ascending order, the accumulation of the sorted daily revenue differences can be obtained. According to the sorted daily revenue difference, 80\% of the annual revenue difference is contributed by the top 19.45\% of the daily revenue difference. This means that the potential profit at node ROCK does not overwhelm the potential profit at node KINCA on every single day. Most of the annual revenue difference is just accumulated in a short period.

%
\vspace{-0.4cm}
\begin{figure}[H]
	\centering
	\includegraphics[width=3.0 in]{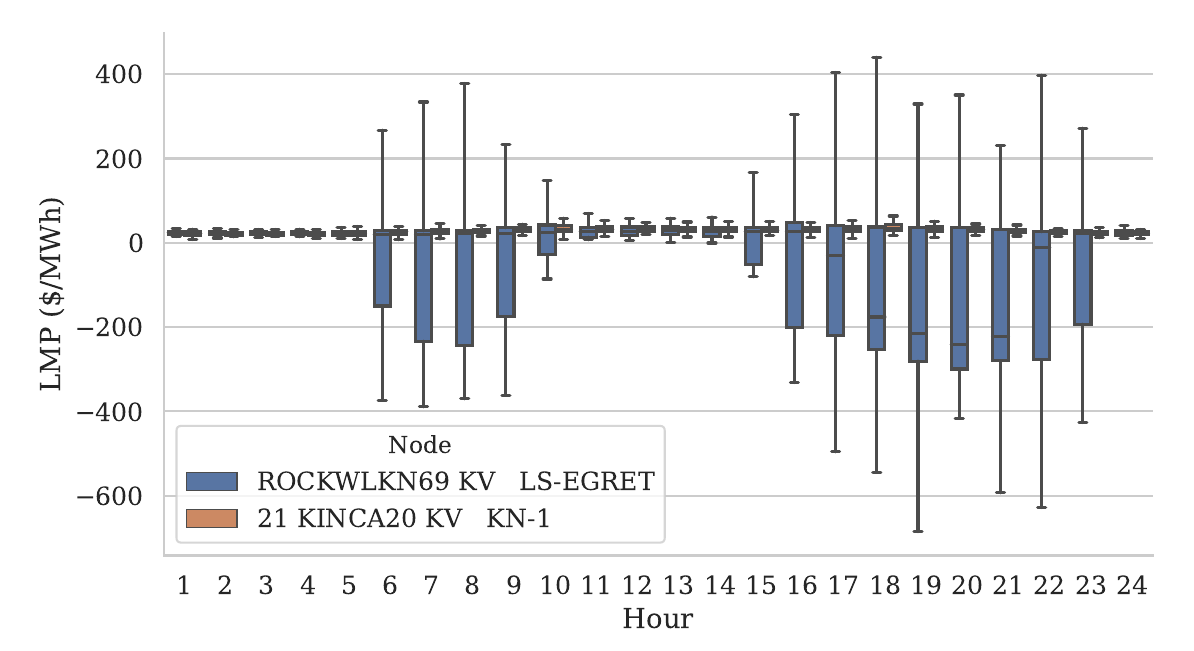}\vspace{-0.15cm}
	\caption{LMP features of the top 20\% daily revenue difference.}
	\label{80Percent_Revenue_LMPs}
\end{figure}

\vspace{-0.4cm}
%
%
Furthermore, to obtain more details about the revenue difference, the hourly LMPs of KINCA and ROCK nodes on the days of the top 19.45\% daily revenue differences are extracted. The boxplots of the extracted 24-hour LMPs in Fig. \ref{80Percent_Revenue_LMPs} shows the hourly LMPs at ROCK have much higher volatility than the LMPs at KINCA. At node ROCK, the LMPs from 6:00 a.m. to 9:00 a.m., and from 1:00 p.m. to 11:00 p.m., are distributed on a larger price range than the other hours, which enables the BESS to have more arbitrage opportunities in the energy market. Notably, a large amount of negative prices is found at node ROCK, and the negative rates make the charging action profitable.\par
\vspace{-0.5cm}
 \vspace{-0.1cm}\par
\begin{figure}[H]
	\centering
	\includegraphics[width=2.8 in]{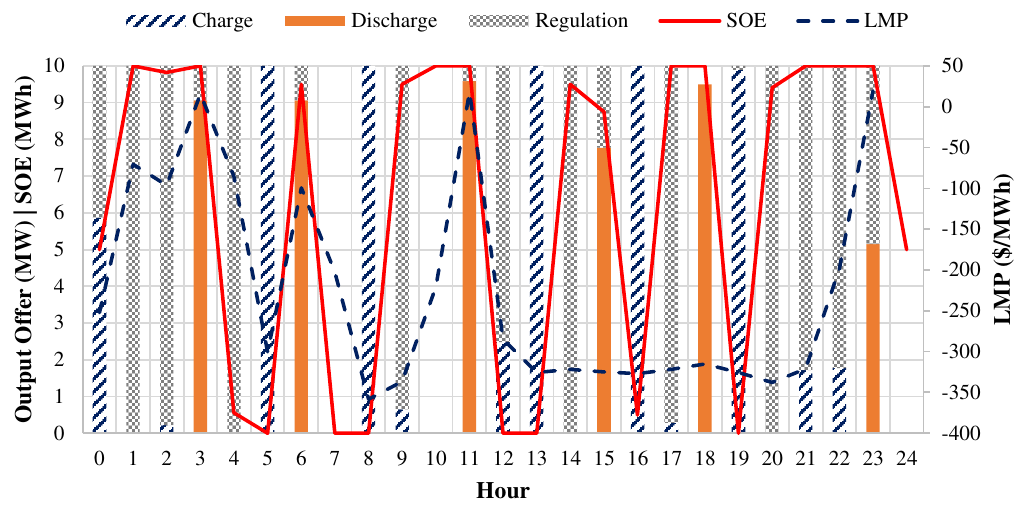}\vspace{-0.cm}
	\caption{Output actions of node ROCK on 9/27/2018.}
	\label{Actions_BestNode}
\end{figure}
\begin{figure}[H]
	\centering
	\includegraphics[width=3.0 in]{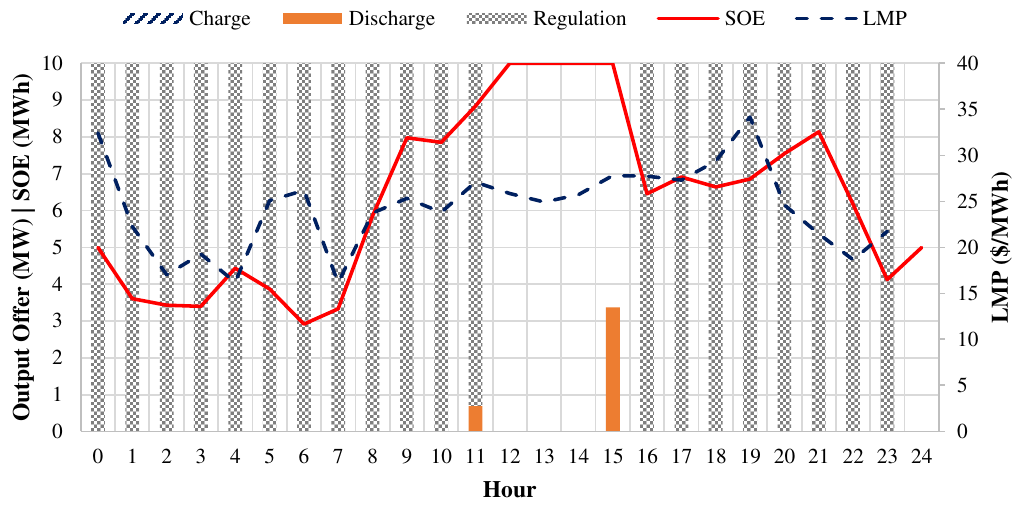} \vspace{-0.1cm}
	\caption{Output actions of node KINCA on 9/27/2018.}
	\label{Actions_WorstNode}
\end{figure}

According to the optimization results on 9/27/2018, the battery at node ROCK can earn \$17k while the battery at node KINCA can just earn \$5.9k. There is \$11k revenue difference, which is more than 189\% of the daily revenue at node KINCA on 9/27/2018. This is the most significant difference in daily revenue between these two nodes in 2018. To better illustrate the BESS behavior, their optimized generation outputs on 9/27/2018 are extracted and displayed in Figs. \ref{Actions_BestNode} and \ref{Actions_WorstNode}. Given the optimized outputs, their hourly SOCs on 9/27/2018 can be calculated by \eqref{MPRE_eq4}. As a reference, the corresponding hourly state of energy (SOE) derived from SOC is plotted with the LMPs in Figs. \ref{Actions_BestNode} and \ref{Actions_WorstNode}.
\vspace{-0.1cm}
 \vspace{-0.1cm}
In Figs. \ref{Actions_BestNode} and \ref{Actions_WorstNode}, the left y-axis indicates the hourly output power and the SOE of the BESS. The y-axis on the right shows the LMP at the node. The bars of different colors at each hour display charging power, discharging power, and regulation output, respectively. The solid and dash lines show the hourly SOEs and LMPs in 24 hours. \par
As shown in Fig. \ref{Actions_BestNode}, the LMP at node ROCK is very volatile, and the peak prices on 9/27/2018 happen at 3:00 a.m., 6:00 a.m., 11:00 a.m., and 11:00 p.m. when the discharging actions are taken for arbitraging in the energy market. When the LMPs are low, the charging actions are found at 5:00 a.m., 8:00 a.m., 1:00 p.m., and 7:00 p.m. \par
\vspace{-0.1cm}
 \vspace{-0.1cm}
In Fig. \ref{Actions_WorstNode}, the standard deviation of 24-hour LMPs is \$4.67/MWh. Compared to the LMPs in Fig. \ref{Actions_BestNode} with \$128.11/MWh standard deviation, the LMPs at KINCA is less volatile, which indicates fewer arbitrage opportunities. Besides, the degradation cost, implemented in the market participation models, forces the BESS not to charge or discharge any single MWh when the arbitrage revenue cannot cover the degradation cost. Due to the degradation cost, it can be seen in Fig. \ref{Actions_WorstNode} that most of the output power at node KINCA is offered to the regulation market. From 12:00 to 15:00, the BESS at KINCA does not even provide full output capacity since the earnings from energy and regulation markets during these hours are not able to cover the degradation cost. \par
Comparing the discharging actions at node KINCA to the multiple charging and discharging actions at node ROCK, the factor that leads to an \$11k daily revenue difference on 9/27/2018 can be discovered. It is the volatility of LMP at node ROCK, which makes the BESS produce more revenue. As mentioned earlier, the LMP is a combination of three components. Since there is only one energy price for all the buses in the system and the energy loss is relatively small compared to the other components \cite{kirschen2018fundamentals}, the power system congestion becomes the dominating factor that influences the potential revenue for the BESS, participating in the energy and regulation markets. In other words, to achieve more revenue, the BESS should search for a place having a highly volatile congestion component. \par
An interesting point that can be observed in Fig. \ref{Actions_WorstNode} is the advantage of participating in both the energy and regulation markets. According to the optimization results, the battery at node KINCA can still make \$5.9k profit on 9/27/2018, which means the BESS can always guarantee a certain amount of profit by participating in the regulation market even though there is a lack of arbitrage opportunity in the energy market. \par
%
%
\vspace{-0.1cm}
\subsection{Potential Revenue of Transportable BESS}
\vspace{-0.1cm}
%
Recalling from the previous section, a significant part of the annual revenue difference is accumulated in a short period. Meanwhile, the renewable energy resources have a strong seasonal pattern \cite{mulder2014implications}. Thus, if the BESS is on a transportable platform (such as trucks) that can be installed and commissioned at different places in different seasons or months, it can capture more revenue in the market and better accommodate to the renewable energy resources and transmission congestion. A transportable BESS can also be formed by aggregating a fleet of EVs that have the V2G capabilities. To investigate the potential revenue of the transportable BESS, the cases of settling the BESS at 4 Locations in 4 Seasons (4L4S), and 12 Locations in 12 Months (12L12M) are studied in the following. \par
Based on the annual motor carrier operations cost report \cite{ATRI_report} provided by American Transportation Research Institute (ATRI), the average carrier cost, including driver costs, fuel, insurance, permits, tolls, etc., is set to \$66.65/hour. Similar to \cite{he2018spatiotemporal}, the size of BESS on each truck is assumed to be 2.5MW/2.5MWh. Therefore, the 10MW/10MWh BESS analyzed can be carried by four trucks. The maximum transportation distance in PJM is expected to be 900 miles, which can cover the distance from Chicago, IL, to Virginia Beach, VA. The average speed of the truck is assumed to be 60 miles/hour, so the maximum transportation time is about 15 hours and transportation cost is \$4k. In addition to the cost during transportation, it is assumed that the wage of a professional electrician is \$65/hour and the integration work will take 10 electricians 4 hours to complete. Then the labor cost for integration is estimated at \$3k, including the electrician cost, insurance, etc. Meanwhile, according to the PJM transmission tariff \cite{PJM_Manual_14A,PJM_Open_Access_Transmission_Tariff}, the cost for transmission interconnection of a 10MW/10MWh BESS is assumed to be \$2.2k, which includes the costs for application, feasibility study, and non-refundable deposit. Hence, the maximum cost for each re-location of the 10MW/10MWh BESS is \$9.2k. It should be noted that the above re-location cost does not apply to a virtual transportable BESS, consisting of aggregated EVs. \par
\begin{figure}[H]
	\centering
	\includegraphics[width=3.2 in]{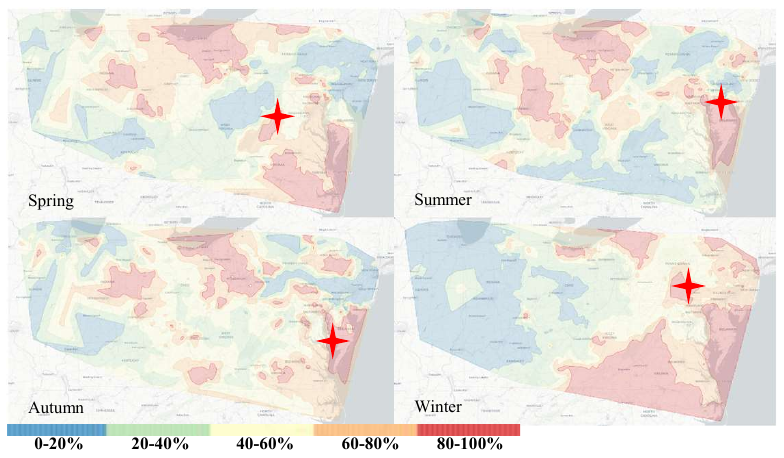}
	\caption{Revenue percentiles by season.}
	\label{PJM_Transportable_Seasons} \vspace{-0.2cm}
\end{figure}

%
According to the optimization results, all nodes’ daily revenues can be obtained and aggregated to the seasonal revenues. Similar to the analysis of annual revenue percentiles, seasonal revenue percentiles are grouped into five intervals and mapped to five different colors. As shown in Fig. \ref{PJM_Transportable_Seasons}, the seasonal revenue percentiles are plotted by spring, summer, fall, and winter, correspondingly. According to the contour maps in different seasons, it can be observed that nodes having the top 20\% revenue on different areas reveal a positive sign for the transportable BESS. The sign indicates the portable BESS can earn more profit by moving from one comparably profitable place to another one in different seasons. In other words, based on the seasonal revenue analysis, the mobile BESS can always chase a better profitable node through the entire market in different seasons to achieve a higher potential of earning. \par
\vspace{-0.3cm}
 \par
More specifically, the most profitable nodes and their seasonal earnings are obtained based on the daily maximum revenues, as listed in Table \ref{tb_Seasonal_Revenue}. As a comparison, the seasonal revenues of 1L1Y are shown in Table \ref{tb_Seasonal_Revenue}. The seasonal profits of 4L4S surpass the seasonal profits of 1L1Y by \$28k, \$33k, and \$118k in spring, summer, and winter, respectively. In fall, the nodes having the best potential revenue of 1L1Y and 4L4S are the same. Therefore, if the BESS is transportable, the best possible earning for 4L4S is \$2.9M, which is \$180k higher than the annual revenue at the best fixed location (i.e. node ROCK). Even considering the \$37k transportation cost for 4L4S, the transportable BESS is still more profitable than stationary BESS in 1L1Y. According to the price node mapping by state, zip code, and transmission zone \cite{PJM_LMP_Model}, these nodes are marked on the contour maps as shown in Fig. \ref{PJM_Transportable_Seasons}. It can be seen that all the nodes with the best seasonal revenues are in the eastern region of PJM. In addition to the more profitable chances, these nearby nodes can lower the moving and relocation cost for mobile BESSs. \par
\vspace{-0.1cm}
\begin{table}[H]
	\centering
	\caption{Seasonal Revenues of 1L1Y and 4L4S}\label{tb_Seasonal_Revenue}
	\begin{tabular}{c|cc||cc}
		\hline \textbf{Season}	& \textbf{Node}  & \textbf{1L1Y Profit} & \textbf{Node}    & \textbf{4L4S Profit}\\ \hline
		\textbf{Spring} & ROCK & \$688k & STRASBUR & \$716k \\ \hline
		\textbf{Summer} & ROCK & \$489k & CEDARCRE & \$523k \\ \hline
		\textbf{Autumn} & ROCK & \$769k & ROCK & \$769k \\ \hline
		\textbf{Winter} & ROCK & \$837k & GARDNERS & \$955k \\ \hline
	\end{tabular} 
\end{table}
The 12L12M scheme is studied for transportable BESS as well. In the 12L12M case, the BESS are transported every month to maximize the revenue in the energy and regulation markets. According to the monthly revenue listed in Table \ref{tb_Monthly_Revenue}, the potential revenue of 12L12M is \$3.2M. Even considering with \$110k transportation cost in a year, the maximum profit of 12L12M is \$177k higher than the revenue of 4L4S and \$319k more than the revenue of 1L1Y. \par
\vspace{-0.1cm}
\begin{table}[H]
	\centering
	\caption{Revenues of 12L12M}\label{tb_Monthly_Revenue}
	\begin{tabular}{c|c|c|c||c|c|c}
		\hline \textbf{Season}	& \multicolumn{3}{c||}{\textbf{Spring}}  & \multicolumn{3}{c}{\textbf{Summer}} \\ \hline
		\textbf{Month} & 3 & 4 & 5 & 6 & 7 &8 \\ \hline
		\textbf{Profit} & \$191k & \$262k & \$313k & \$185k & \$179k & \$205k \\ \hline

		\hline\hline \textbf{Season} & \multicolumn{3}{c||}{\textbf{Autumn}}    & \multicolumn{3}{c}{\textbf{Winter}}\\ \hline
		\textbf{Month} & 9 & 10 & 11 & 12 & 1 & 2 \\ \hline
		\textbf{Profit} & \$218k & \$317k & \$234k & \$220k & \$726k & \$163k \\ \hline
	\end{tabular} 
\end{table}

\section{Optimal Placement of BESSs} \label{RevenuePrediction}
%
The previous revenue analysis is based on historical data. However, the most profitable nodes identified using historical data may not be the appropriate locations for BESSs in the future. Hence, an optimization algorithm is needed for placing and operating the BESSs. This algorithm is particularly important for the transportable BESSs since they can move to different locations for more profitable opportunities. To achieve this goal, an optimal placement algorithm is proposed in this section. Without loss of generality, the algorithm that aims at the optimal placement for the next month can be extended to other future times, such as next season or year.
%
%
%
\subsection{Prediction of LMP's Volatility}
%
Recalling from the previous section, the revenue difference among the nodes is caused by the volatility of LMP in the energy market. Hence, to identify the most profitable nodes, the prediction of LMP volatility becomes a critical task. Thus, a standard ARIMA model \cite{box2015time}, which is effective for forecasting price sequences \cite{conejo2005day, zhao2017improving, zhao2018improvement}, is used. An ARIMA model can be described by \eqref{ARIMA_Model} \cite{box2015time}.
\begin{equation} \label{ARIMA_Model} 
{\phi }_p\left(B\right){\mathrm{\nabla }}^dy_t=\mu +{\theta }_q\left(B\right){\varepsilon }_t 
\end{equation}
where $B$ is the backward shift operator; $p$ is the auto-regression (AR) order, which determines how many past values are used for regression; \textit{d} is the differencing order, which represents the number of differencing transformations to make the training series stationary; \textit{q} is the moving-average (MA) order to determine how many previous error terms ${\varepsilon}_t$ should be considered. The error terms ${\varepsilon}_t$ are independent and identically distributed noise with zero mean and finite variance. \par
%
%
Given the hourly LMP, the monthly standard deviations of LMP at each node can be calculated. For formulating a stationary series for the ARIMA model, the time series $y_{i,t}$ is obtained by a natural logarithm transformation of a differential series, as shown in \eqref{Std_Difference}.
\begin{equation} \label{Std_Difference}
y_{i,t} = log(\sigma_{i,t}-\sigma_{PJM,t}+c)
\end{equation}
where $\sigma_{i,t}$ is the standard deviation of LMP at node $i$ in month $t$; $\sigma_{PJM,t}$ represents the standard deviation of system's LMP in month $t$; c is a positive constant offset to guarantee the logarithm transformation; and, $y_{i,t}$ is the transformed time series in month $t$, which is used by the ARIMA model to forecast the LMP volatility at node $i$. The LMPs from 1/1/2016 to 12/31/2018 of each node are used for training the ARIMA model to predict the corresponding standard deviation in the next month. The market data between 1/1/2019 and 6/30/2019 are used to validate the proposed algorithm. 
\vspace{-0.1cm}
\subsection{Clustering of Pricing Nodes}
\vspace{-0.1cm}
In addition to the estimation of LMPs' volatility, the clustering of nodes is important for the optimal BESS placement. After grouping the nodes according to the LMPs' features in the past month, the algorithm can place the transportable BESSs at the nodes in different clusters for the purpose of reducing the risk and seeking for more profitable opportunities throughout the market. The K-means \cite{liao2005clustering}, one of the most widely used clustering algorithms, is implemented to cluster the nodes into different groups via minimizing the objective function described in \eqref{k_means_obj}:
\begin{equation}
Min(\sum_{k=1}^{K}\sum_{x \in G_k}^{} {\left\| x-\mu_k\right\|}^2) \label{k_means_obj}
\end{equation}
where $K$ is the total number of clusters, which is determined by the elbow method \cite{kodinariya2013review}; $G_k$ represents the $k$-th group of the clustered data $x$; $\mu_k$ is the mean of $x$ in $G_k$ (also called centroid). The objective of K-means is to minimize the sum of the Euclidean distance between the data points and the centroid of the corresponding group. \par
To address the high-dimensional hourly LMP for K-means clustering, the hourly LMP in the previous month is grouped by each day to obtain the daily standard deviation. Given the daily standard deviations of each node, the Principle Component Analysis (PCA) \cite{ding2004k} is implemented to further reduce the dimension of the historical data. The PCA is a linear transformation to represent the original data by a set of orthogonal variables. Via PCA, the original data can be described with a low dimensional subspace with keeping much of the variance in the dataset. $x$ in \eqref{k_means_obj} is the PCA components of normalized daily standard deviation of the LMPs at each node. By implementing the PCA and K-means for the LMPs in the previous month, the clusters of different nodes can be obtained.
%
%
\vspace{-0.1cm}
\subsection{Optimal Placement Algorithm}
%
Given the prediction of LMP's monthly standard deviations and the clusters of nodes, an algorithm shown in \eqref{opt_pla_obj}-\eqref{opt_pla_Geo_LB} is designed to assist the market participants for optimal placement of BESSs. The objective is to maximize the sum of the predicted LMP's volatility of the selected nodes, as shown in \eqref{opt_pla_obj}.
\begin{gather}
Max(\sum_{i=1}^{N}\delta_i \: \sigma_{i,t+1}') \label{opt_pla_obj}
\end{gather}
\begin{center}\vspace{-0.1cm} s.t. \eqref{opt_pla_Nes}-\eqref{opt_pla_Geo_LB}. \vspace{-0.15cm}\end{center}
\begin{gather}
\sum_{i=1}^{N}\delta_i = {N}_{ES} \label{opt_pla_Nes}\\
\sum_{i \in G_k}^{}\delta_i \le \overline{N}_{Clu} \label{opt_pla_maxN_cluster}\\
\delta_i \: \delta_j \: d_{i,j}^{Geo} \ge \underline{D}^{Geo} , \; \;  \forall i < j \label{opt_pla_Geo_LB}
\end{gather}
where $\sigma_{i,t+1}'$ is the forecast monthly standard deviation of the LMPs at node $i$; $\delta_i$ is a binary decision variable. When $\delta_i=1$, the node $i$ is selected for placing transportable BESS. Equation \eqref{opt_pla_Nes} indicates that the algorithm needs to find $N_{ES}$ locations for placing the BESSs. Equation \eqref{opt_pla_maxN_cluster} enforces that no more than $\overline{N}_{Clu}$ BESSs can be placed in cluster $G_k$. Since the nodes in the same cluster could have similar characteristics of LMPs, \eqref{opt_pla_maxN_cluster} enables the market participants to place the BESSs in different clusters for the risk diversification. Equation \eqref{opt_pla_Geo_LB} enforces that the geographical distance $d_{i,j}^{Geo}$ between nodes $i$ and $j$ must be larger than $\underline{D}^{Geo}$. This information is beneficial for the risk management of market participants. When more system information is available to the market participants, the geographical distance can be extended to other types of distance, such as the electrical distance between the nodes.
Furthermore, \eqref{opt_pla_Geo_LB} can be transformed to linear constraints as shown in \eqref{opt_pla_Geo_LB_rlx1}-\eqref{opt_pla_Geo_LB_rlx3}, where $\delta_{i,j}$ is a binary variable and $M$ is a large positive number. 
Therefore, the optimal placement algorithm can be solved as the following mixed-integer linear programming problem. \vspace{-0.05cm}
\begin{gather}
\delta_{i,j} \ge \delta_i+\delta_j-1, \; \; \forall i < j \label{opt_pla_Geo_LB_rlx1}\\
\delta_{i,j} \le \delta_i, \; \delta_{i,j} \le \delta_j, \; \; \forall i < j \label{opt_pla_Geo_LB_rlx2}\\
\delta_{i,j} \: d_{i,j}^{Geo}+(1-\delta_{i,j}) \: M \ge \underline{D}^{Geo} , \; \; \forall i < j \label{opt_pla_Geo_LB_rlx3}
\end{gather} 
%
\subsection{Performance of the Proposed Algorithm}
In order to evaluate the algorithm's performance, we assume that the market participant needs to place 5 10MW/10MWh BESSs in the market for the next month, which indicates $N_{ES}=5$. Meanwhile, $\overline{N}_{Clu}$ is set to 1, which means no more than 1 BESS can be placed in the same cluster of nodes. The minimum distance $\underline{D}^{Geo}$ between two BESSs is chosen to be 50 miles. With the implementation of the proposed optimal placement algorithm for the first 6 months in 2019, the optimal nodes for placing the BESSs in each month can be obtained. For comparison, a base case for placing the BESSs in the market based on the historical revenue is included, where the market participants use the 5 most profitable nodes in the current month as the placements for the next month. The market data, such as RTLMP, RMCCP, RMPCP, and the regulation signals, from 1/1/2019 to 6/31/2019, are collected to verify the algorithm. \par
\vspace{-0.3cm}
\begin{table}[H]
	\centering
	\caption{Performances of the Base Case and the Proposed Algorithm}\label{tb_Optiaml_Algorithm_Revenue}
	\begin{tabular}{c|cc|c|cc}
		\hline \textbf{Mo.}	& \textbf{Base}  & \textbf{Proposed}  & \textbf{Mo.}	& \textbf{Base}  & \textbf{Proposed}\\ \hline
		\textbf{1} & \$551k & \$623k  &	\textbf{4} & \$560k & \$618k \\ \hline
		\textbf{2} & \$513k & \$559k  &	\textbf{5} & \$434k & \$509k \\ \hline
		\textbf{3} & \$673k & \$733k  &	\textbf{6} & \$592k & \$475k \\ \hline
	\end{tabular} \vspace{-0.3cm}
\end{table}
As shown in Table \ref{tb_Optiaml_Algorithm_Revenue}, the optimal placement algorithm can outperform the base case in 5 out of 6 months. In the first 5 months, the algorithm can earn 11.4\% more revenue than the base case. Even though the base case beats the proposed algorithm by \$117k in the last month, the optimal placement algorithm can still make \$194k more profit than the base case. In other words, the optimal placement algorithm is capable of benefiting market participant by identifying the potentially profitable sites to deploy the BESSs. \par
%
\section{Conclusion} \label{conclusion}
%
This paper carried out a comprehensive and system-wide analysis of potential revenues of BESSs in a market-based power system. The PJM market has been used as a case study, and the features of PJM market data of different locations and seasons have been characterized for the analyses. Based on the geographic distributions of BESS revenue, a set of candidate sites have been identified for installing a BESS in the market to achieve the highest revenue. By comparing the best and worst profitable nodes, LMP volatility has been found to be the dominating factor for revenue maximization. Most of the annual revenue difference was observed to be accumulated in a short period, which reveals the potential opportunity of transportable BESS for capturing more profits from different locations in several periods. Through the comprehensive study of different seasons and months at all the nodes in the whole system, the results indicate a transportable BESS is capable of gaining a higher revenue than a corresponding stationary BESS in the energy and regulation markets. Furthermore, an algorithm has been developed and verified for optimal placement of BESSs. The results are encouraging to the BESS market participants for investment and management in the energy and frequency regulation markets. Meanwhile, the increasing number of BESSs can help power systems accommodate to more sustainable energy.
\par
\bibliographystyle{IEEEtran}
\bibliography{references}

%








\end{document}